\documentclass[aps,amsfonts,amsmath,prd,preprint,nofootinbib,longbibliography,tightenlines]{revtex4-1}
\newcommand{\beq}{\begin{equation}}
\newcommand{\eeq}{\end{equation}}
\usepackage{graphicx}

\def\bea{\begin{eqnarray}}
\def\eea{\end{eqnarray}}
\def\bml{\begin{subequations}}
\def\blea{\bml\bea}
\def\eml{\end{subequations}}
\def\elea{\eea\eml}
\def\Im{\mathop{\rm Im}}
\def\sign{\mathop{\rm sign}}

\def\bn{\mathbf{n}}
\def\bk{\mathbf{k}}
\def\bx{\mathbf{x}}
\def\by{\mathbf{y}}
\def\bz{\mathbf{z}}
\def\bX{\mathbf{X}}
\def\nsf{\mathsf{n}}
\def\nmax{n_{\text{max}}}
\def\ohat{\hat{\mathbf{\Omega}}}
\def\Re{\mathop{\mathrm{Re}}}
\def\Im{\mathop{\mathrm{Im}}}
\def\bI{\mathbf{I}}
\def\ogw{\Omega_{\text{gw}}}
\def\rhogw{\rho_{\text{gw}}}
\def\Pgw{P_{\text{gw}}}
\def\teq{t_{\text{eq}}}
\def\tep{t_{\text{ep}}}

\def\year{\text{yr}}

\def\Gpc{\text{Gpc}}
\def\Hz{\text{Hz}}

\begin{document}

\title{Stochastic gravitational wave background from smoothed cosmic string loops}

\author{Jose J. Blanco-Pillado}
\email{josejuan.blanco@ehu.es}
\affiliation{Department of Theoretical Physics, University of the Basque Country, Bilbao, Spain}
\affiliation{IKERBASQUE, Basque Foundation for Science, 48011, Bilbao, Spain}

\author{Ken D. Olum}
\email{kdo@cosmos.phy.tufts.edu}
\affiliation{Institute of Cosmology, Department of Physics and Astronomy,
Tufts University, Medford, MA 02155, USA}

\def\changenote#1{\footnote{\textbf{#1}}}


\begin{abstract} 

We do a complete calculation of the stochastic gravitational wave
background to be expected from cosmic strings.  We start from a
population of string loops taken from simulations, smooth these by
Lorentzian convolution as a model of gravitational back reaction,
calculate the average spectrum of gravitational waves emitted by the
string population at any given time, and propagate it through a
standard model cosmology to find the stochastic background today.  We
take into account all known effects, including changes in the number
of cosmological relativistic degrees of freedom at early times and the
possibility that some energy is in rare bursts that we might never
have observed.

\end{abstract}
 
\maketitle
\section{Introduction}

Our universe may contain a network of cosmic strings, which could be
either flux tubes arising from a symmetry-breaking transition at high
energies or the fundamental strings of superstring theory (or
one-dimensional D-branes) stretched out to astrophysical lengths 
\cite{Vilenkin:2000jqa,Dvali:2003zj,Copeland:2003bj}.  The
best way to discover such a network, if it exists, is to observe the
stochastic background of gravitational waves emitted by cosmic string
loops.  Non-observation of such a background in pulsar timing
arrays currently gives the strongest bounds on the energy scale
of a possible cosmic string network.

In usual models, cosmic strings do not have ends.  Thus they exist as
a ``network'' of infinite strings and closed loops.  Intersections
between strings lead to reconnections, and when a string intersects
itself, it produces a loop.  Loops then oscillate relativistically and
decay by the emission of gravitational waves.  If the string energy
per unit length is $\mu$, the gravitational power emitted is $\Gamma G
\mu$, where $\Gamma$ is a number of order 50 depending on the
shape of the string, and $G$ is Newton's constant.  We work in units
where $c=1$.

In both the matter and the radiation era, the flow of energy from long
strings into loops and thence into gravitational waves maintains the
network in a scaling regime, where all linear measures, such as the
average distance between strings, stay at a fixed multiple of the
horizon distance (or the age of the universe).  Scaling allows us to
extrapolate over many orders of magnitude between what can be studied
in a simulation and the universe today.

To connect observations or observational limits to the properties of
possible cosmic strings, we need to accurately compute the spectrum of
gravitational waves to be expected from a cosmic string network of a
given energy scale.  The steps in this process are as follows.
\begin{enumerate}

\item\label{item:simulation} First we simulate the network of cosmic strings to find the rate
  of production of loops of various sizes from the long string
  network, and we extract a representative sample of loop shapes from
  the simulation.
\item\label{item:toy} This gives the distribution of loop shapes at the time the loops
  are formed, but gravitational back reaction modifies these shapes.
  Since we do not yet have a code for calculating these changes in
  shape, we use a toy model of smoothing to estimate them.  
\item\label{item:spectrum} We then compute the gravitational spectrum and total power
  $\Gamma$ for each loop.
\item\label{item:density} Using $\Gamma$, which also gives the
  evaporation rate, we integrate the production and evaporation
  processes over cosmological time to determine the distribution of loops
  existing at each redshift $z$.
\item\label{item:allspectrum} We integrate the spectrum of individual
  loops over the loop distribution at each $z$ to find the overall
  emission spectrum.
\item\label{item:background} Then we integrate the emission spectrum
  over cosmological time to get the present-day background.
\end{enumerate}

Items \ref{item:simulation}, \ref{item:toy}, and \ref{item:density}
have already been done in
Refs.~\cite{BlancoPillado:2011dq,Blanco-Pillado:2015ana,Blanco-Pillado:2013qja}.
The purpose of this paper is to complete the program with items
\ref{item:spectrum}, \ref{item:allspectrum}, and
\ref{item:background}.  We include all known effects except that we
use a smoothing model rather than computing directly the effects of
gravitational back reaction on loop shapes.  A companion paper
compares the results with current observations \cite{Blanco-Pillado:2017rnf}.

It is traditional in papers such as this to consider ``small loop''
models in which the predominant size of loops at production is $\Gamma
G \mu$ times the production time, so loops last for only about one
Hubble time.  In our opinion, there is no reason to consider such
models any more.  They were inspired by early simulations
\cite{Bennett:1987vf,Allen:1990tv} that found loops at the resolution
scale, but recent simulations
\cite{Vanchurin:2005yb,Ringeval:2005kr,Vanchurin:2005pa,Martins:2005es,Olum:2006ix,BlancoPillado:2011dq},
with much greater reach, found loop production at scales related to
the horizon size at the time of production.

Refs.~\cite{Hindmarsh:2008dw, Bevis:2010gj,Hindmarsh:2017qff}
simulated cosmic string networks in lattice field theory, rather than
treating strings as linelike objects as in
Refs.~\cite{Albrecht:1984xv,Bennett:1987vf,Allen:1990tv,
  Vanchurin:2005yb,Ringeval:2005kr,Vanchurin:2005pa,Martins:2005es,Olum:2006ix,BlancoPillado:2011dq}.
The results were radically different.  Long strings were subject to
strong damping and almost no oscillating loops were found in these
simulations.  However, we cannot imagine how such results could be
applicable to the astrophysical situation.  The present-day ratio of
loop size or curvature scale, $\Gamma G \mu t_0$, to the string
thickness, about $\sqrt{\mu/\hbar}$, is $\Gamma (G \mu)^{(3/2)}
t_0/t_{\text{Planck}} \sim 10^{44}$.  Thus on any possible scale
relevant to field theory dynamics, strings are straight to
fantastically good approximation, and their motion should be given by
the Nambu-Goto equations of motion. Indeed this was shown to be the
case in simulations of individual Abelian-Higgs strings when the
curvature scale was larger than the thickness
\cite{Olum:1998ag,Moore:1998gp,Olum:1999sg}.

The remainder of this paper is structured as follows.  In the next
section we calculate the gravitational wave background in terms of
the expansion history of the universe, the distribution of loops at
each epoch, and the power spectrum of gravitational waves emitted from
a typical loop.  We discuss these three components in turn in
Secs.~\ref{sec:cosmology}, \ref{sec:n}, and \ref{sec:Pn}.
Sec.~\ref{sec:results} gives our results, and we conclude in
Sec.~\ref{sec:conclusion}.

Some technical matters are deferred to appendices.
Appendix~\ref{sec:cusps} gives the details of the calculation of the
radiated power from a cusp. Appendix~\ref{sec:maxn} discusses how many harmonics need to be
computed to find the spectrum in any given direction. Appendix~\ref{sec:summation} discusses
summing contributions from the discrete modes emitted by loops, and 
Appendix~\ref{sec:bursts} considers whether the fact that some power
is in very rare bursts requires a modification of the stochastic
background calculation.

For previous estimates of the stochastic spectrum of gravitational
waves from cosmic strings with various assumptions see
Refs.~\cite{Vilenkin:1981bx,Hogan:1984is,Vachaspati:1984gt,Accetta:1988bg,
  Bennett:1990ry,Caldwell:1991jj,Siemens:2006yp,DePies:2007bm,Olmez:2010bi,Sanidas:2012ee,Sanidas:2012tf,Binetruy:2012ze,Kuroyanagi:2012wm,
  Kuroyanagi:2012jf,Blanco-Pillado:2013qja}. For a discussion of burst signals
  from cosmic strings see \cite{Damour:2000wa,Damour:2001bk,Damour:2004kw,Siemens:2006vk}.

\section{Stochastic gravitational wave
  background}\label{sec:background}

We will compute the stochastic background of gravitational waves
presently existing as the fraction of the critical density given by
the energy of gravitational waves in unit logarithmic interval of
frequency,\footnote{The same background can be expressed as its power
  spectral density,
\beq
S_h(f) = \frac{3 H_0^2}{2\pi^2 f^3} \ogw(\ln f)\,,
\eeq
or as its characteristic strain $h_c = \sqrt{f S_h(f)}$.  Pulsar
timing arrays use these quantities directly, but interferometers
adjust them by averaging interferometer sensitivity over
polarization and arrival direction, reducing $S_h$ by 5 and $h_c$ by
$\sqrt5$ over the values given here.  For a clear explanation of
various measures of the background see
Ref.~\cite{Moore:2014lga}.}
\beq\label{eqn:Omega}
\ogw (\ln f) = \frac{8\pi G}{3H_0^2} f \rhogw(t_0,f)\,,
\eeq
where $\rhogw$ is the energy density in gravitational waves per unit
frequency.  Since gravitational waves persist from very early times,
the energy in a comoving region is just the redshifted total energy deposited
there,
\beq
\rhogw (t_0,f) = \int_0^{t_0}{ \frac{dt}{(1+z(t))^4} \Pgw(t,f')
\frac{\partial f'} {\partial f}}\,,
\eeq
where $\Pgw(t,f')$ is the total gravitational wave power of all loops
existing at time $t$ into unit range of emitted frequencies.  The
emitted frequency that becomes frequency $f$ today is just $f'
=(1+z)f$, so we find
\beq
\rhogw (t_0,f) = \int_0^{t_0}{\frac{dt}{(1+z(t))^3}
   \Pgw \left(t,(1+z)f\right)}\,.
\eeq

Now let $\nsf(l,t)$ be the density of loops per unit volume per unit
range of loop length $l$ existing at time $t$.  We will model the
emission from these loops as given by some power spectrum $P_n$,
giving the power in  harmonic $n$ (frequency $2n/l$) in units of
$G\mu^2$.  For a given $n$, loops with a range of lengths $dl$
will emit in a frequency range $df' = - (f'/l)dl$, and thus
\beq
\Pgw (t,f') = G \mu^2\sum_{n=1}^{\infty} \frac{l}{f'} \nsf(l,t) P_n\,,
\eeq
with $l=2n/f'$.  Thus we can write
\beq\label{eqn:rhocp}
\rhogw (t,f) =G \mu^2 \sum_{n=1}^{\infty} C_n P_n\,,
\eeq
with 
\beq
C_n (f) = \int_0^{t_0}\frac{dt}{(1+z)^5} \frac{2n}{f^2} \nsf(l,t) \,.
\eeq
We can now change the integration variable using
\beq\label{eqn:dt}
dt = -\frac{dz}{H(z) (1+z)}
\eeq
to get
\beq\label{eqn:Cn}
C_n(f) = \frac{2n}{f^2 }\int_0^{\infty} \frac{dz}{H(z) (1+z)^6}
 ~\nsf\left(\frac{2n} {(1+z)f},t(z)\right)
\eeq
Equations~(\ref{eqn:rhocp},\ref{eqn:Cn}) give the stochastic
background in terms of the cosmology $(H(z), t(z))$, the loop density $\nsf(l,t) $, and the
radiation power spectrum of each loop, $P_n$.  In the following, we will discuss these effects
in turn.

\section{Cosmology}\label{sec:cosmology}

The cosmological dependence in Eq.~(\ref{eqn:Cn}) is in $H(z)$ in the
denominator and $t(z)$ appearing as an argument to the loop
distribution $\nsf(l,t) $.  We will consider a flat
$\text{radiation}+\text{matter}+\Lambda$ cosmology, with
\beq
H(z) = H_0 \sqrt{\Omega_{\Lambda} + (1+ z)^3\Omega_m +  G(z)(1+z)^4\Omega_r}\,,
\eeq
where the function
\beq\label{eqn:G}
G(z) =\frac{T(z)^4g_*(z)}{T_0^4 (1+z)^4g_{*,0}}
\eeq
corrects for the change in the number of relativistic degrees of
freedom at early times.  Here $T(z)$ is the temperature at redshift
$z$, $g_*(z)$ the effective number of relativistic degrees of freedom
then, and $T_0$ and $g_{*,0}$ these quantities today.

Neutrinos today are presumably nonrelativistic and should technically
be included in $\Omega_m$ and not $\Omega_r$.  But the value of
$\Omega_r$ is important only at early times when neutrinos were
relativistic.  So we define $\Omega_r$ here to be the value it would
have with massless neutrinos of temperature (because neutrino
decoupling takes place before electron-positron annihilation)
$(4/11)^{1/3}$ times the present cosmic microwave background
temperature.

The age of the universe at redshift $z$ is the integral of Eq.~(\ref{eqn:dt}),
\beq
t(z)= \int_z^\infty\frac{dz'}{H(z') (1+z')}\,,
\eeq
where we will use \cite{Ade:2015xua}
\blea
\Omega_\Lambda &=& 0.69\,, \\
\Omega_m &=& 0.31\,,
\elea
and we can compute
\beq
\Omega_r =\frac{32\pi G\sigma g_{*,0}T_0^4}{3H_0^2}\,,
\eeq
where $\sigma$ is the Stefan-Boltzmann constant, and
$g_{*,0}\approx 3.36$ is the effective number of relativistic degrees
of freedom with photons and massless neutrinos.  With $T_0 = 2.2725K$
and writing $H_0 = 100 h$ km/s/Mpc as usual, we find
\beq\label{eqn:Omegar}
h^2\Omega_r = 4.15 \times 10^{-5}\,.
\eeq
When necessary we will use the value $h=0.68$ \cite{Ade:2015xua}.

\section{Loop density}\label{sec:n}

\subsection{Uniform radiation era}

Reference~\cite{Blanco-Pillado:2013qja} gives the number density of loops in
the radiation era,
\beq\label{eqn:nr}
\nsf_r(l,t) =  \frac{0.18}{t^{3/2} \left(l + \Gamma G \mu t\right)^{5/2}}
\eeq
for $l < 0.1t$.  Equation~(\ref{eqn:nr}) applies when the universe has
been in the radiation era (without changes in the degrees of freedom)
for a long time.  Accordingly it exhibits scaling behavior in which
\beq\label{eqn:nscaling}
\nsf_r(l,t) = t^{-4} \nsf(x)\,,
\eeq
where $x=l/t$ and $\nsf(x)$ is the number of loops per unit $x$ in volume
$t^3$,
\beq
\nsf(x) =  \frac{0.18}{ \left(x + \Gamma G \mu \right)^{5/2}}\,.
\eeq

In this case, there is a simple result.  Deep in the radiation era,
and ignoring changes in the degrees of freedom,
\beq\label{eqn:Hr}
H(z) = (1+z)^2H_r\,,
\eeq
\beq\label{eqn:tr}
t(z) = \frac1{2(1+z)^2H_r}\,,
\eeq
where
\beq
H_r = H_0\sqrt{\Omega_r}
\eeq
is the contribution from radiation to the Hubble constant today.

Putting Eq.~(\ref{eqn:Hr}) into Eq.~(\ref{eqn:Cn}) gives
\beq
C_n (f) = \frac{2n}{f^2} \int \frac{dz}{(1+z)^8H_r} 
 ~\nsf\left(\frac{2n} {(1+z)f},t(z)\right)\,.
\eeq
We use Eq.~(\ref{eqn:nscaling}) to change from $\nsf(l,t)$ to $\nsf(x)$,
with $t$ given by Eq.~(\ref{eqn:tr}), to get
\beq
C_n(f) = \frac{32H_r^3}{f^2} \int dz\,\nsf(x)\,.
\eeq
Then we change the variable of integration from $z$ to
\beq
x= \frac lt=\frac{8n(1+z)H_r}{f}\,,
\eeq
giving
\beq\label{eqn:Cnr}
C_n(f)= \frac{8H_r^2}{f}\int dx\,\nsf(x)\,.
\eeq

From Eq.~(\ref{eqn:Cnr}), we see that $C_n$ has no dependence on $n$,
so the stochastic background depends only on
\beq\label{eqn:GammaPn}
\Gamma =\sum_{n=1}^{\infty} P_n\,,
\eeq
that $C_n$ depends only on the total loop number
density\footnote{This agrees with Eq.~(21) of
  Ref.~\cite{Blanco-Pillado:2013qja}, which gives the number in volume
  $d_h^3 = 8 t^3$}
in volume $t^3$,
\beq\label{eqn:nrtotal}
\int_0^{\infty}{dx\,\nsf(x) } = 0.12 (\Gamma G \mu)^{-3/2}\,,
\eeq
and finally that $C_n(f)\sim 1/f$, so the power per unit logarithmic
interval of frequency, $\ogw(\ln f)$, is constant.

Using
Eqs.~(\ref{eqn:Omega},\ref{eqn:rhocp},\ref{eqn:Cnr},\ref{eqn:GammaPn},\ref{eqn:nrtotal}),
we find
\beq
\ogw(\ln f) = 8.0 \Omega_r\sqrt{\frac{G\mu}\Gamma}\,,
\eeq
and Eq.~(\ref{eqn:Omegar}) gives
\beq\label{eqn:Omegagwr}
h^2 \ogw(\ln f) = 3.3\times 10^{-4}\sqrt{\frac{G\mu}\Gamma}
= 4.7 \times 10^{-5}\sqrt{G\mu}\,,
\eeq
with $\Gamma = 50$.

The high-frequency background comes almost entirely from deep in the
radiation era, so one might expect a plateau in $\ogw(\ln f)$ given by
Eq.~(\ref{eqn:Omegagwr}).  However, we will see below that changes in
the number of degrees of freedom introduce a few smooth steps
on this plateau region of the spectrum.

\subsection{Changes in the number of degrees of freedom}

At early times, the expansion rate of the universe changes because of
the annihilation of relativistic species, which injects additional
energy into the universe and reduces its rate of cooling.  These
changes are incorporated into the function $G(z)$ of
Eq.~(\ref{eqn:G}).  We handle them as follows.

We do not consider changes to the scaling properties of the string
network, but assume that it always traces the current rate of
expansion.  But we do take into account the fact that the important
loops at any given time are relics of earlier times when $G(z)$ was
different.

First consider a universe which spends a long time in a radiation era
with $G(z) = G_1$.  Instead of Eqs.~(\ref{eqn:Hr},\ref{eqn:tr}), we
have
\beq\label{eqn:HrG}
H(z) = (1+z)^2H_r\sqrt{G_1}
\eeq
and
\beq\label{eqn:trG}
t(z) = \frac1{2(1+z)^2H_r\sqrt{G_1}}\,.
\eeq
The loop density is still given by Eq.~(\ref{eqn:nr}), with $t$ from
Eq.~(\ref{eqn:trG}),
\beq\label{eqn:nrG}
\nsf_r(l,t) =  \frac{0.18\cdot 2^{3/2} ( H_r^2 G_1)^{3/4}(1+z)^3}
{\left(l + \Gamma G \mu t\right)^{5/2}}
\eeq
and
\beq\label{eqn:lnzf}
l=\frac{2n}{(1+z)f}\,.
\eeq
We will use Eq.~(\ref{eqn:nrG}) even in the case
where $G(z)$ is changing, although this is not entirely accurate.
Taking into account more effects would lead to an even smoother
dependence of $\ogw$ on $f$.

Once loops of a certain size are no longer being produced in
significant numbers, their density merely dilutes, going as $(1+z)^3$,
and the loops become shorter by gravitational back reaction.  Both
these processes are included in Eq.~(\ref{eqn:nrG}), so
Eq.~(\ref{eqn:nrG}) holds for in any later era, with $G_1$ always
being the $G(z)$ at the time at which the loop was produced.

Of course not all loops of the same size were produced at the same
time, but the loop production function is peaked in a fairly narrow
range of loop size to production time ratio around $0.1$
\cite{Blanco-Pillado:2013qja}, so we will make this approximation to
compute $G_1$.  Suppose a loop with length $l$ at time $t$ was
produced at time $t_1$.  Then its length at production was $l+\Gamma
G\mu (t-t_1)$ and its ratio of length to production time was
\beq
x_1 = \frac{l+\Gamma G\mu (t-t_1)}{t_1}\,.
\eeq
Setting $x_1 = 0.1$ and using the approximation $\Gamma G\mu\ll x_1$, we find
\beq\label{eqn:t1}
t_1 \approx 10(l+\Gamma G\mu t)\,.
\eeq
and then
\beq\label{eqn:G1t1}
G_1 = G(z(t_1))\,.
\eeq
We compute $G(z)$ using a code for $g_*(z)$ written by Masaki Yamada,
which includes the contributions from all the particles in the
Standard Model.  The result is to introduce small steps in the $G(z)$
function, noticeable especially around electron-positron annihilation
and the QCD phase transition.

Putting Eq.~(\ref{eqn:nrG}) in Eq.~(\ref{eqn:Cn}) gives
\beq\label{eqn:CnG}
C_n(f) = \frac{0.18\cdot 2^{5/2}H_r^{3/2}n}{f^2}
\int dz\frac{G_1(l,t)^{3/4}}{H(z)(1+z)^3\left(l + \Gamma G \mu t\right)^{5/2}}\,,
\eeq
where $z$ is the redshift at which the gravitational wave is emitted,
$t=t(z)$ is the age of the universe at redshift $z$, given by
Eq.~(\ref{eqn:tr}) in the radiation era after electron-positron
annihilation, $l$ is given by by Eq.~(\ref{eqn:lnzf}), and $G_1(l,t)$
is computed using Eqs.~(\ref{eqn:t1},\ref{eqn:G1t1}).

\subsection{Matter era}

In the matter era, there are two kinds of loops.  For $G\mu$
compatible with observational bounds, the most important loops were
formed in the radiation era.  Their density is given by
Eq.~(\ref{eqn:nrG}), which already includes dilution as the universe
expands and loss of length due to gravitational radiation.  We thus
use Eq.~(\ref{eqn:CnG}) in all eras, with $H(z)$ and $t(z)$ as
appropriate.

With $t$ in the matter era and $G\mu$ around current limits, we can
set $G_1=1$, although we do not make this approximation in our
numerical calculations.  The largest loops formed at the time of
electron-positron annihilation have size about $0.1 \tep$.  For them
to survive until matter-radiation equality at $\teq$ requires $\Gamma
G \mu \teq<0.1\tep$ and thus $\Gamma G \mu < 0.1 \tep/\teq \approx 6
\times 10^{-13}$, or $G\mu \lesssim 10^{-14}$.

There are also loops formed in the matter era.  Analysis of
simulations \cite{Blanco-Pillado:2013qja} gives the density of such
loops in a scaling regime,
\beq\label{eqn:nm}
\nsf_m(l,t) = \frac{0.27 - 0.45 (l/t)^{0.31}} {t^2 \left(l + \Gamma G \mu t\right)^2}
\eeq
for $l < 0.18 t$.  Using Eq.~(\ref{eqn:nm}) in
Eqs.~(\ref{eqn:rhocp},\ref{eqn:Cn}) gives the stochastic background
arising from these loops.  We give the result in
Sec.~\ref{sec:results}, but it is negligible compared to
the background from relic loops from the radiation era.  The basic reason is that for the
dominant loop size $l\sim\Gamma G \mu t$, Eq.~(\ref{eqn:nr})
is larger than Eq.~(\ref{eqn:nm}) by factor $(\Gamma G \mu)^{-1/2}$.
For $G \mu < 2\times 10^{-11}$, this is at least $2 \times 10^5$.

We did not study loops formed during the matter to radiation
transition.  But these also have little consequence.  In fact, even
loops formed near the end of the radiation era make little
contribution.  From Eqs.~(\ref{eqn:trG},\ref{eqn:t1}) and taking
$t_0H_0\approx1$, $G_1=1$, and using Eq.~(\ref{eqn:Omegar}), we find
the dominant loops today were produced at redshift about
\beq
0.16\Omega_r^{-1/4}(\Gamma G\mu)^{-1/2}\approx 16(\Gamma G\mu)^{-1/2}
> 5 \times 10^5
\eeq
for $G\mu < 2\times 10^{-11}$.  This is far larger than the redshift of
matter-radiation equality, about 3000.

\section{Spectrum of a loop}\label{sec:Pn}

\subsection{Population of loops}

The last ingredient is $P_n$, the average gravitational spectrum radiated
from a loop.  We compute this separately for loops formed in the matter
era and those formed in the radiation era.  In each case, we use a
sample of loops found in simulations. (See
Ref.~\cite{BlancoPillado:2011dq} for a discussion of simulation
techniques.).  We used $1060$ loops in the radiation era and $812$ in the
matter era.

These simulation loops, however, are not representative of loops
existing at any given time, because those loops have lost a
significant fraction of their energy due to gravitational wave
emission and thus have had their shapes modified by back reaction.
For the present paper, we model this effect by smoothing the loops by
convolving them with a Lorentzian \cite{Blanco-Pillado:2015ana}, even
though we know \cite{Wachter:2016hgi,Wachter:2016rwc} that this model
is not entirely correct.  We consider the last three smoothing steps,
corresponding to loss of 1/8, 1/4, and 1/2 of the initial loop length.
In the next section we give some separate results for these three steps,
but for the final result we used only the last step.  Including the
others would not make any noticeable difference.

Convolution yields a set of smooth loops whose radiation power $P_n$
we would like to compute.  We should not model these loops in a
piecewise linear form, as we do for loops in our simulations.  A
piecewise linear loop would have kinks between the pieces, and at
sufficiently high frequencies these fictitious kinks could make a big
difference to the gravitational radiation power.

Instead, we represent the strings as smooth functions given by their
Fourier transforms.  We keep the Fourier amplitudes for some finite
number $N_f$ (up to 4096) of frequencies.  To compute the
gravitational radiation spectrum of such loops, we must understand
their motion, which we now discuss.

\subsection{Loop motion and cusps}

The expansion of the universe is very important for the evolution of
the string network and later for the propagation of gravitational
waves.  But the loops we will study are always much smaller than the
Hubble distance, and so their evolution takes place essentially in
flat space.

The general solution for the motion of a Nambu-Goto string in flat
spacetime can be written
\beq
X^\mu(t,\sigma) = {\frac12} \left[X^\mu_-  (\sigma_-) + X^\mu_+ (\sigma_+)\right]\,,
\eeq
where $\sigma_\pm = t\pm \sigma$, are the lightcone coordinates on the
string worldsheet built from the timelike coordinate $t$ and the
spacelike parameter $\sigma$. We will work in the gauge where the
4-vector functions $X^{\mu}_\pm$ have $X^0_\pm = \sigma_\pm$, and the
spatial part obeys the constraints $|\bX'_- (\sigma_-)| = |\bX'_+
(\sigma_+)| = 1$, where, as usual, the prime denotes a derivative of
the function with respect to its argument.  The two functions
$\bX_\pm$ specify the motion of the loop.  It is these functions that
we smooth to emulate gravitational back reaction effects, and it is
these smooth functions that we represent by their Fourier
coefficients.

For a closed loop in the rest frame, $\bX_\pm$ are periodic,
$\bX_\pm (\sigma_\pm) = \bX_\pm (\sigma_\pm + l)$, and thus
\beq
\int_0^l\bX'_\pm (\sigma_\pm) d \sigma_\pm= 0\,.
\eeq
Thus $\bX'_+$ and $\bX'_-$ each trace out a loop on the
``Kibble-Turok'' unit sphere
\cite{Kibble:1982cb,Turok:1984cn,Garfinkle:1987yw}, and the center of
gravity of the loop is at the center of the sphere.  Generically these
two paths will cross, so there are usually points where
\beq
\bX'_+ (\sigma^c_+) = \bX'_- (\sigma^c_-) \,.
\eeq
Thus at $t_c = (\sigma_+^c+\sigma_-^c)/2$, $\sigma_c =
(\sigma_+^c-\sigma_-^c)/2$, the string velocity (formally) reaches the
speed of light,
\beq
\left|{{d\bX}\over{dt}}\right| =1\,,
\eeq
and the string doubles back on itself,
\beq
{{d\bX}\over{d\sigma}} =0\,,
\eeq
so such a point is called a cusp.\footnote{Note that cusps are not
  artifacts of Nambu-Goto dynamics.   In fact they are formed in
  field theory cosmic strings, as we showed \cite{Olum:1998ag}
  in the Abelian-Higgs model.}

The existence of cusps leads to difficulties in computing the
gravitational radiation spectrum from a loop.  When there is a cusp,
the spectrum falls only as $n^{-4/3}$ \cite{Vachaspati:1984gt}, where $n$ is the
harmonic number of the radiation.  Thus the integrated power falls
only as $n^{-1/3}$.  This slow decrease makes it impractical to
accurate compute the total power by simply computing numerically up
to some maximum $n$.  Instead we compute the power from cusps
analytically (See Appendix~\ref{sec:cusps}), and use this computation for high frequencies in
directions near cusps.

Cosmic strings may also have kinks: places where there is a
discontinuous change in $\bX'$.  These lead to a spectrum which falls
as $n^{-5/3}$ \cite{Garfinkle:1987yw}.  However, in the present analysis, kinks are
smoothed out by convolution, so that we do not have to consider them
in our computations.  A better analysis of kink evolution
\cite{Wachter:2016hgi} show that kinks are opened out rather than
being rounded off.  In future work we will compute the actual back
reaction numerically, but at the moment we are restricted to modeling
it as a smoothing process.

\subsection{Radiation power}

Computation of the radiation power spectrum, $P_n$, for each of our 
loops proceeds as follows.  First we find
cusps, the places where the paths of $\bX'_+$ and $\bX'_-$ cross on
the unit sphere.  We do this by generating by fast Fourier transform
(FFT) at least $10 N_f$ samples of each function and looking for
crossings between the great-circle paths connecting adjacent samples.
When we find such a crossing, we narrow it down using the Fourier
transform representations of $\bX_\pm$.

Then we integrate the gravitational radiation power over solid angle
by dividing the sphere of emission directions into triangles.  We
start with an icosahedron projected onto the sphere and then
repeatedly divide each triangle into 4 smaller triangles by inserting
a point at the center of each edge \cite{Atkinson:sphere}.  If we
perform the division process $N_{\text{split}}$ times, the total
number of triangles is $20\times 4^{N_{\text{split}}}$.  We used
$N_{\text{split}}=5$.

We now see how close each triangle comes to the direction (i.e., the
$\bX'_+= \bX'_-$) of any cusp.  If there is a cusp inside the triangle
or within a threshold angle, taken as 0.1, we compute the emission
using the cusp emission procedure described Sec.~\ref{subsec:cusp}
below.  If not, we compute the radiation using the generic expression
for the power given in Sec.~\ref{subsec:generic} in the direction of
the center of the triangle (given by the normalized average of the 3
corner directions) and multiply by the area of the triangle on the
unit sphere.

\subsection{Radiation in a generic direction}
\label{subsec:generic}

To compute the radiation power in a given direction $\ohat$, we follow
Refs.~\cite{Burden:1985md,Allen:1991bk,Allen:2000ia}.  The angular
power density emitted in harmonic $n$ is
\beq\label{eqn:dPdO}
\frac{dP_n}{d\Omega}
= \frac{G\mu^2l^2}{2\pi} \omega_n^2 (|A_+|^2+|A_\times|^2)
= 8 \pi G\mu^2  n^2 (|A_+|^2+|A_\times|^2)\,,
\eeq
where $l$ is the length of the loop, $\omega=4\pi n/l$ and $A_+$ and $A_\times$ are the amplitudes of
the two gravitational wave polarizations.  If we construct a
coordinate system whose $z$ axis is in the $\ohat$ direction, they are
given by
\bml\label{eqn:A+x}\bea
A_+ &=& I^-_xI^+_x-I^-_yI^+_y\,,\\
 A_\times &=& I^-_yI^+_x+I^-_xI^+_y\,,
\elea
where
\beq\label{eqn:I}
\bI^{\pm (n)}(\ohat) = \frac1l\int_0^l d\sigma_\pm\, \bX'_\pm(\sigma_\pm)
e^{(2\pi i n/l)( \sigma_\pm - X_z(\sigma_\pm))}\,.
\eeq

From Eqs.~(\ref{eqn:A+x}) we find
\bea
|A_+|^2 &=& |I^-_x|^2|I^+_x|^2+|I^-_y|^2|I^+_y|^2-2 \Re(I^-_xI^{-*}_yI^+_xI^{+*}_y)\,,\\
|A_\times|^2 &=& |I^-_y|^2|I^+_x|^2+|I^-_x|^2|I^+_y|^2+2 \Re(I^{-*}_xI^-_yI^+_xI^{+*}_y)\,,
\eea
where asterisk means complex conjugation.  Thus
\beq\label{eqn:AA}
|A_+|^2+|A_\times|^2 = |I^-_\perp|^2|I^+_\perp|^2
+ 4 \Im(I^-_xI^{-*}_y)\Im(I^+_xI^{+*}_y)\,,
\eeq
where $|I^\pm_\perp|^2 = |I^\pm_x|^2+|I^\pm_y|^2$.
We can write
\beq
\Im(I^\pm_xI^{\pm*}_y) = (\bI_I \times \bI_R)_z\,,
\eeq
where the subscripts $I$ and $R$ mean the imaginary and real parts of
the vector.  This shows that the result is independent of the choice
of the coordinate system in the perpendicular plane.

We would now like to compute $\bI^{\pm(n)}_\perp$ in directions far
from any cusp, for specific $\bX_\pm$ given in terms of their Fourier
transforms.  To do this quickly, we would like to use FFT to compute
all necessary $n$ at once.  However, Eq.~(\ref{eqn:I}) does not have
the form of a Fourier transform, because the exponent is not simply
$2\pi i n \sigma_\pm/l$.  But we can approximate it as a nonuniform
discrete Fourier transform as follows.

First take $N$ positions $\sigma_j=jL/N$, $j=0\ldots N-1$.  To compute
$I^+_x$, for example, we generate $X'_x(\sigma_j)$ and $\phi_j = (\sigma_j -
X^+_z(\sigma_j))/l$ at these $N$ positions.  This can be done by FFT
using the Fourier components of $X^+_x$.  We then have
\beq
I_x^{+(n)}(\ohat) = \frac1N\sum_{j = 0}^{N-1} X'^+_x(\sigma_j)
e^{2\pi i n \phi_j}\,.
\eeq
This is a non-uniform Fourier transform problem, which can be solved
in $O(N\ln N)$ time.  We use the method of Potts, Steidl, and Tasche
\cite{nufft}.  The choice of how many $n$ to compute is discussed in
Appendix~\ref{sec:maxn}.

\subsection{Radiation in a cusp direction}
\label{subsec:cusp}

In the case where the triangle is close to the direction of the cusp,
the situation is more difficult.  In any given direction the
gravitational power from the cusp decreases with frequency only as
$\omega^{-2/3}$, so the power per logarithmic interval of $\omega$
increases as $\omega^{1/3}$.  This continues until the radiation is
cut off at some maximum frequency proportional to $\theta^{-3}$, where
$\theta$ is the angle between the cusp direction and the direction of
observation.  The angular area over which a given frequency $\omega$
is important is proportional to $\theta^2\sim\omega^{-2/3}$, so the
radiation from a cusp, integrated over solid angle, declines as
$\omega^{-4/3}$ and the contribution per logarithmic interval goes as
$\omega^{-1/3}$.

This long tail makes it difficult to compute the radiation accurately
using the techniques above.  First, we would need huge numbers of
harmonics near the cusp, and second, the high-frequency radiation
varies rapidly over small distances within the triangle.  To solve
this problem, we calculate the high-frequency cusp radiation
analytically using a simple model of the cusp, and then integrate
numerically over the triangular region.  

Because this model does not
work well for low frequencies, we compute those using
Eqs.~(\ref{eqn:dPdO}-\ref{eqn:I}) even in the direction of the cusps.  Because of aliasing, FFT techniques do not give
accurate answers even at low frequencies, unless all frequencies with
significant power are included.  So we compute the integral in
Eq.~(\ref{eqn:I}) directly.  The decision of which frequencies are
done by which technique is made by using the cusp technique whenever
the frequency would have significant variation over the range of
directions in the triangle.

The details of the cusp procedure are given in Appendix \ref{sec:cusps}.
We show in Fig.~\ref{fig:radiation-distribution}
\begin{figure}
\includegraphics[width=5in]{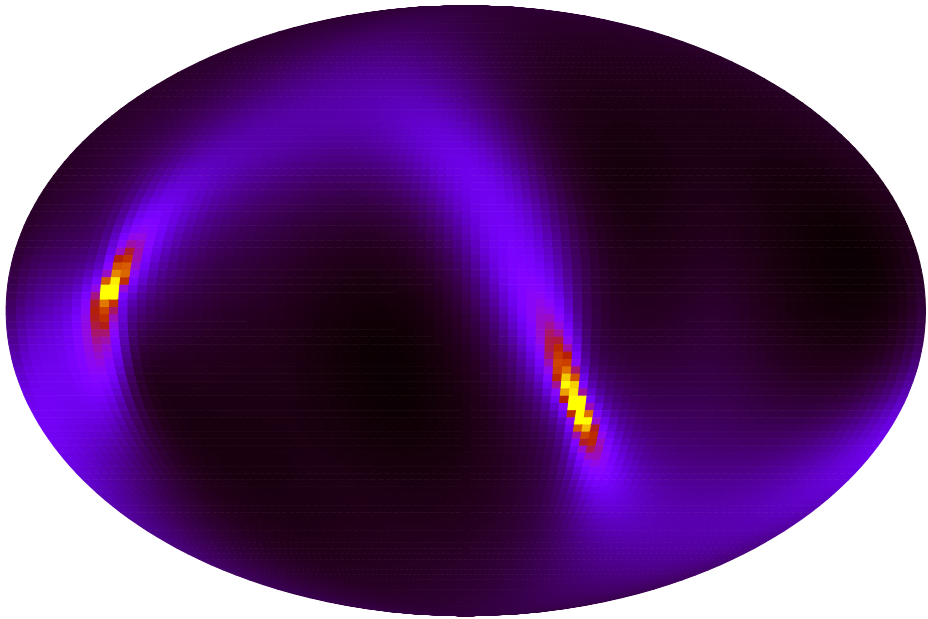}
\caption{Spatial distribution of the gravitational radiation from a smooth loop. We show
the Mollweide projection of the radiation density on a sphere surrounding the loop. 
The brighter regions represent the high radiation density in the direction of the cusps.}
\label{fig:radiation-distribution}
\end{figure}
an example of the 
radiation density emitted by a typical loop. We see the enhancement
of the radiation density along the directions of the cusps.

\section{Results}\label{sec:results}

\subsection{Total radiation power $\Gamma$ of a loop}

The simplest result that one can obtain is the total radiation power,
integrated over directions and frequencies.  This has the form
$P=\Gamma G \mu^2$, so the goal is to determine the constant $\Gamma$.
The slowest known radiator is the Allen-Casper-Ottewill (ACO) loop
\cite{Allen:1994bs} studied extensively by Anderson
\cite{Anderson:2005qu,Anderson:2008wa,Anderson:2009rf}, with $\Gamma
\approx 39.0025$.  There is no upper limit to $\Gamma$.  For example,
the $\Gamma$ of Burden \cite{Burden:1985md} loops grows without bound
as the angle between the planes of $\bX^+$ and $\bX^-$ decreases. The
power spectrum and the total power emitted from these loops can be
computed using the expressions found in \cite{Burden:1985md}. We have
used these simple loop solutions as test beds for our numerical
code. The results are in very good agreement with the analytic
calculations.

\begin{figure}
\includegraphics[width=5.5in]{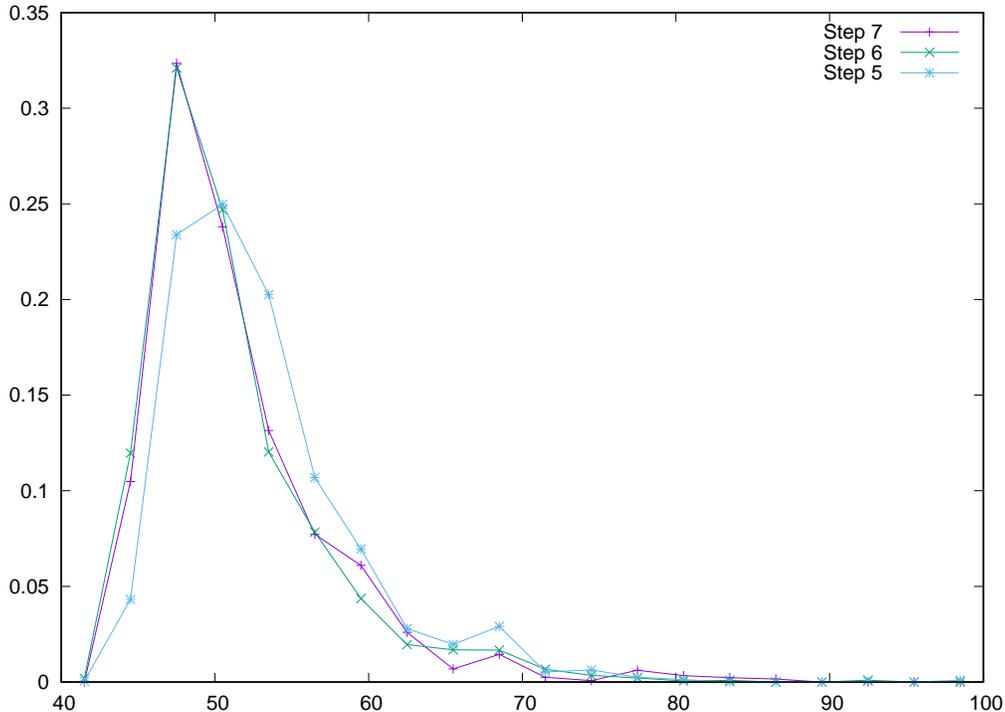}
\caption{Histogram of $\Gamma$ values for the loops in the radiation
  era at various stages of smoothing. The results for the matter era
  are very similar.}
\label{fig:gammas-3-steps}
\end{figure}

\begin{figure}
\includegraphics[width=5.5in]{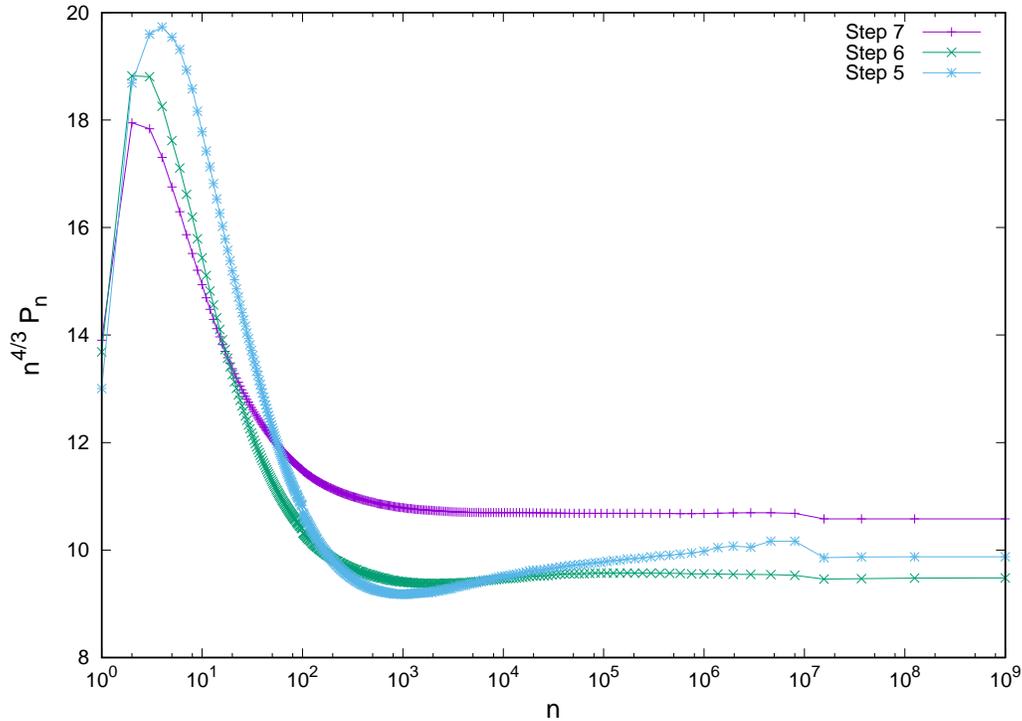}
\caption{Average power spectrum scaled by $n^{4/3}$ of radiation era
  loops at three stages of smoothing.  The feature at
  $n\approx 10^7$ is an artifact.  See the end of Appendix~\ref{sec:maxn}}
\label{fig:PS-several-steps}
\end{figure}

A histogram of $\Gamma$ for loops taken from simulations with various
degrees of smoothing is shown in Fig.~\ref{fig:gammas-3-steps}.
Remarkably, for the great majority of loops, $\Gamma\sim 50$.  Since
smoothing the loop produces cusps that were not there before, one
might think that smoother loops would have higher radiation power.
However, as shown in Fig.~\ref{fig:PS-several-steps},
the additional power emitted by the cusp comes at the expense of
non-cusp emission.  Thus the presence of cusps moves the power to
higher frequencies but produces little change in the overall power.
  
Consequently it is a good approximation to use $\Gamma = 50$ always,
and we do not need to concern ourselves with the fact that different
loops evaporate at different rates. We have calculated the average value of
$\Gamma$ for a population of $1060$ loops obtained in 3 radiation
era runs and obtained $\bar \Gamma_{r}= 51.43$.

In the matter era we consider $812$ loops and the average total radiation
power is $\bar \Gamma_{m}= 53.55$.

\subsection{Power spectrum}

\begin{figure}
\includegraphics[width=5.2in]{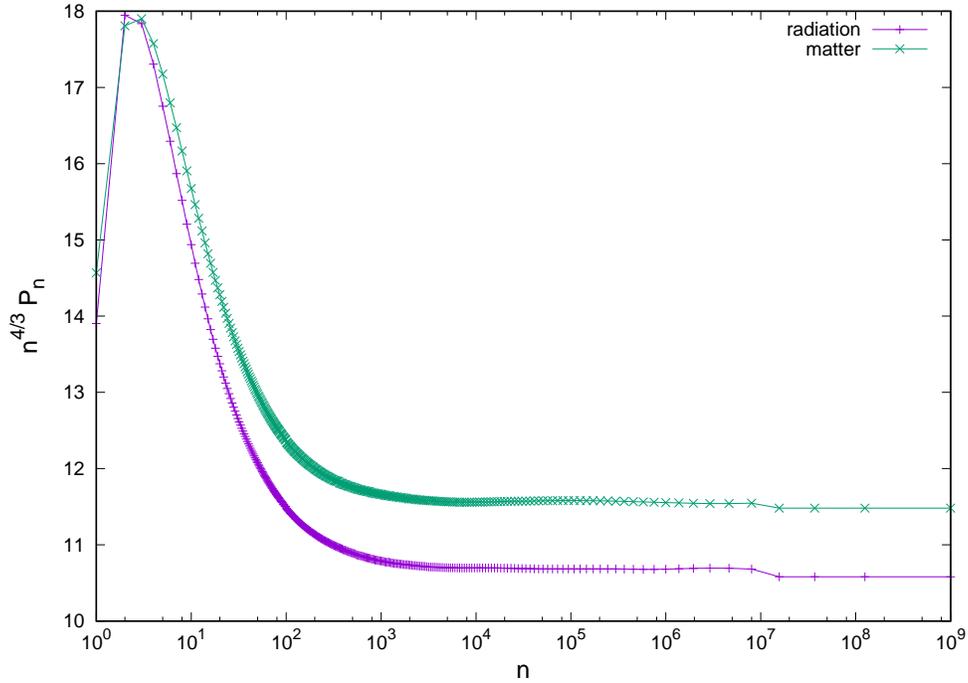}
\caption{Average power spectrum scaled by $n^{4/3}$ of radiation and matter era loops. We show here only the power
spectrum at the last step on the smoothing procedure.  See the end of
Appendix~\ref{sec:maxn} for a discussion of the artifact at $n\approx 10^7$}
\label{fig:PS-rad-mat}
\end{figure}

The power spectrum of the loop is the set of discrete numbers $P_n$,
$n=1\ldots\infty$.  We use this spectrum in Eq.~(\ref{eqn:rhocp}) to
compute $\rhogw(f)$ and so $\ogw(\ln f)$.  But of course we cannot
compute an infinite set of numbers.  Instead we compute a finite
number of $P_n$, with the $n$ chosen to give an accurate result in
$\rhogw(f)$, taking account of our expectation that $P_n$ will drop as
$n^{-4/3}$.  The details are given in Appendix~\ref{sec:summation}.
We take a weighted average\footnote{See Ref.~\cite{Blanco-Pillado:2015ana} for a detailed 
description of the weighting procedure to compute the averages from our
sample of loops from the simulation.} of the $P_n$ of the smoothed loops from the
simulation to use in Eq.~(\ref{eqn:rhocp}).  The average $P_n$ for
loops in the radiation and matter eras are shown in Fig.~\ref{fig:PS-rad-mat}.

\begin{figure}
\includegraphics[width=5.2in]{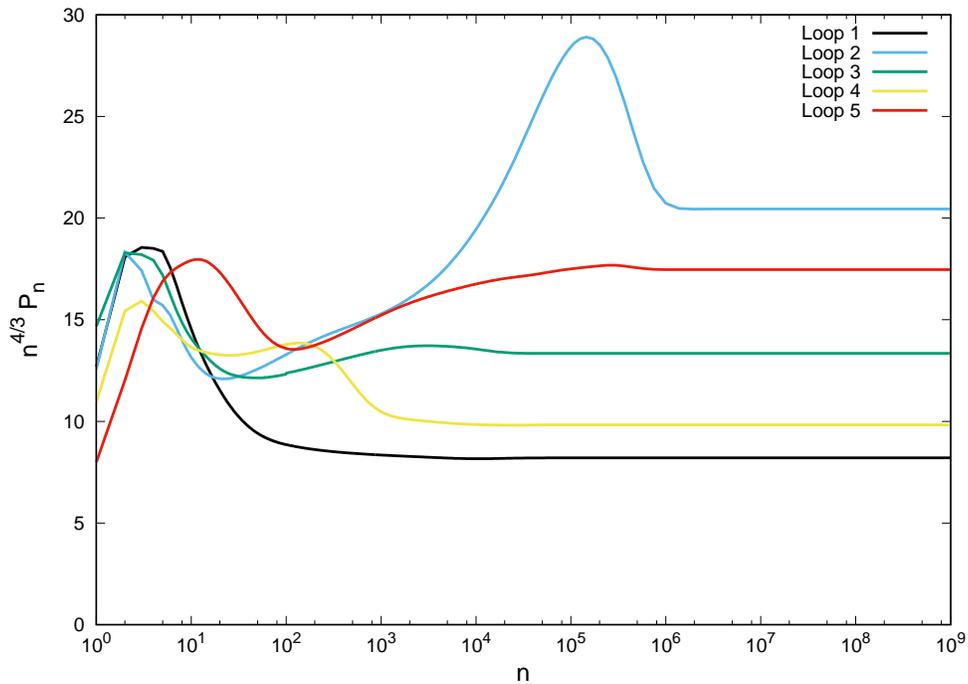}
\caption{Power spectra of a few individual loops in the radiation era,
  chosen to show the diversity of possibilities.  The great majority
  of loops have spectra similar to loop 1 here, but some are quite
  different.  Nevertheless, averaging over many loops gives the smooth
  spectra shown in Fig.~\ref{fig:PS-rad-mat}.}
\label{fig:PS-diversity}
\end{figure}

We note that even though the average power spectrum is very smooth,
some of the loops have quite different shapes, 
which leads to some variety in the power spectra as shown in
Fig.~\ref{fig:PS-diversity}.
 Of course this variation is amplified by the way we choose to represent
the power spectrum by $n^{4/3} P_n$.

\subsection{Stochastic Gravitational Wave Spectrum: $\ogw (\ln f) $}

\begin{figure}
\includegraphics[width=6.5in]{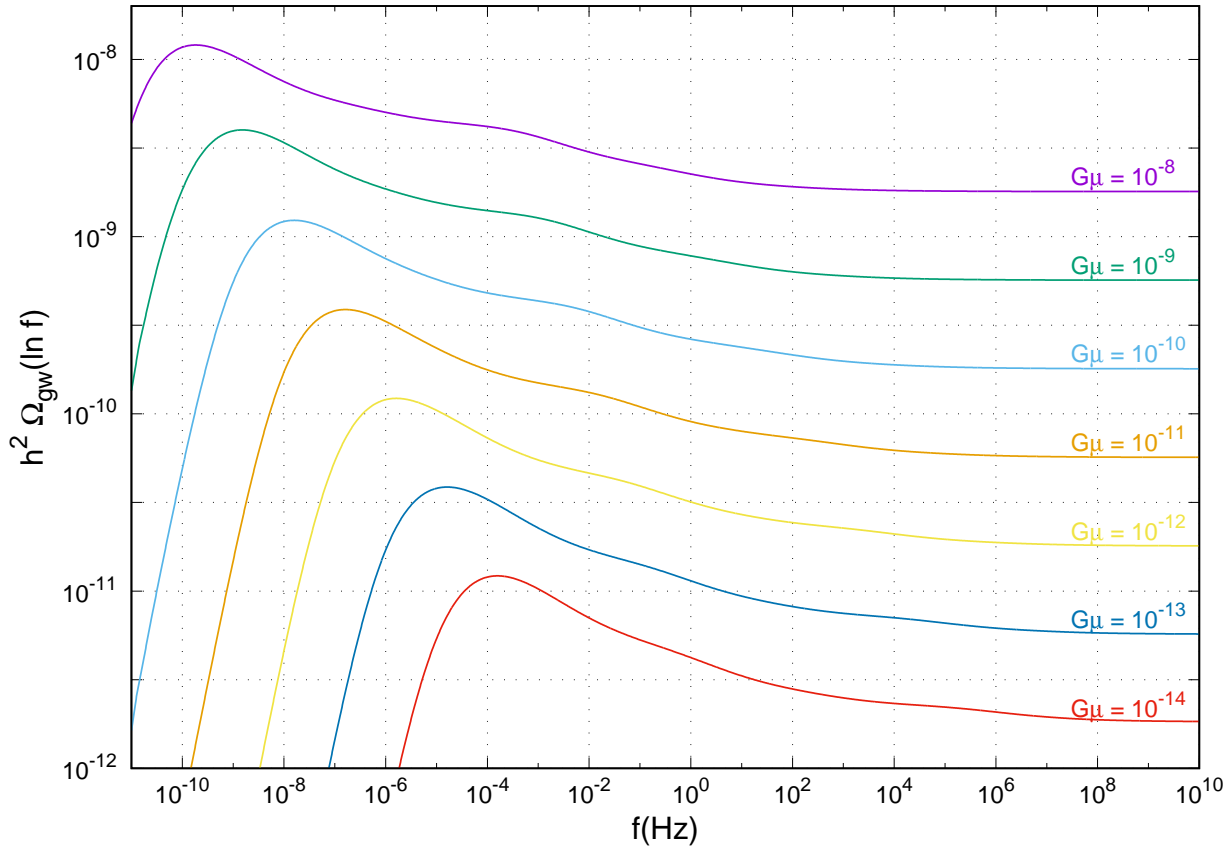}
\caption{The stochastic gravitational wave spectrum for string
  tensions between $G\mu = 10^{-8}$ and $10^{-14}$.}
\label{fig:Omega-f}
\end{figure}

With the $P_n$, and the $C_n$ from Eq.~(\ref{eqn:Cn}) using the loop
densities computed in Sec.~\ref{sec:n}, we compute $\ogw(\ln f)$ for a
range of frequencies $f$ using Eqs.~(\ref{eqn:Omega},\ref{eqn:rhocp}).
The results are shown in Fig.~\ref{fig:Omega-f}.
This includes the contributions from the loops in all the eras, as
described earlier in the text.

The general form of the spectra can be understood as follows.  Very
low frequencies can only be emitted by large loops, but large loops
are suppressed by $l^{-5/2}$, as in Eq.~(\ref{eqn:nr}).  There is an
extra power of $f$ in Eq.~(\ref{eqn:Omega}), so at very low
frequencies, $\ogw$ goes as $f^{-3/2}$.  At even lower frequencies
there is a cutoff because there are essentially no strings of size
$l>0.1t$ at time $t$, but this does not appear in
Fig.~\ref{fig:Omega-f}.

At high frequencies, we are sensitive only to loops radiating in the
radiation era.  According to Eq.~(\ref{eqn:Omegagwr}) this would give
a plateau proportional to $\sqrt{G\mu}$.  However, changes in the
number of relativistic degrees of freedom at early times turn the
plateau into a series of decreasing plateaus, which are smoothed into
a decline with some wiggles.\footnote{Note that adding new ingredients
  in the thermal history of the universe, such as new physics beyond
  the Standard Model, could introduce new features in this
  spectrum. In principle, detecting this stochastic background from
  strings could allow us to probe the thermal history of the universe,
  though in fact the effect occurs only at very high frequencies}  At
intermediate frequencies is there is a peak resulting from
gravitational wave emission in the matter era.

Decreasing $G\mu$ does not change the frequencies at which any given
loop radiates, but the overall power drops proportionately to $G\mu$.
Simultaneously, the lower $G\mu$ allows loops to survive longer, so
that at any given time there is now a new, larger population of older
and smaller loops, which radiate at higher frequencies.  The net
result is that the curve of $\ogw$ moves downward proportionately to
$\sqrt{G\mu}$ and to the right as $1/(G\mu)$.

To model intercommutation probability $p<1$ in the standard
way\footnote{We are skeptical about this procedure.  While increasing
  the loop density by $1/p$ reproduces the $p=1$ average reconnection
  rate between unrelated strings, the production of loops requires a
  long string to intersect with itself, and the chance of that is
  unaffected by the overall density.  So the evolution of a $p<1$
  network may be more complicated than a simple rescaling.} one
should move up the graph for the desired $G\mu$ by factor $1/p$.

\section{Conclusion}\label{sec:conclusion}

We have computed the stochastic background of gravitational waves to
be expected from a network of Abelian-Higgs cosmic strings with $G\mu$
ranging from $10^{-8}$ to $10^{-14}$.  We used a $\Lambda CDM$
cosmology with string loops taken from simulations and smoothed by
Lorentzian convolution as a model of gravitational backreaction
\cite{Blanco-Pillado:2015ana}.  We analyzed strings in the radiation
era, strings from the radiation era radiating in the matter era, and
strings produced in the matter era (though these, and all strings in
the recent $\Lambda$-dominated era, make no essentially no
contribution).  We took account of changes in the number of
relativistic degrees of freedom in the very early universe, which give
an important reduction in the background at high frequencies.  We
found (see Appendix~\ref{sec:bursts}) that there is no need remove
energy contained in rare bursts from the observable stochastic
background.  The only missing ingredient is a real calculation of
gravitational backreaction, which the subject of work currently in
progress.

A companion paper \cite{Blanco-Pillado:2017rnf} compares the results predicted
here with limits from current observations and discusses the prospects
of detection in the future.  The data shown in Fig.~\ref{fig:Omega-f},
with the range $G\mu$ extended down to $10^{-25}$, are available at\\
\url{http://cosmos.phy.tufts.edu/cosmic-string-spectra/}.

\section*{Acknowledgments}

We thank Masaki Yamada for providing us with a code to compute the
$g_*$ function, and Xavier Siemens and Alex Vilenkin for helpful
conversations.  This work was supported in part by the National
Science Foundation under grant numbers 1213888, 1213930, 1518742, and
1520792.  J. J. B.-P.  is supported in part by the Basque Foundation
for Science (IKERBASQUE), the Spanish Ministry MINECO grant
(FPA2015-64041-C2-1P) and Basque Government grant (IT-979-16).

\appendix

\section{Gravitational wave power from cusps}\label{sec:cusps}

In this appendix we compute the gravitational radiation power due to
the string near a cusp in directions close to the direction in which
the cusp is moving, $\bX_+ =\bX_-$.  To simplify the calculation, we
will choose our coordinate system so that the $z$ axis lies in this
direction (note that this is a different convention from that of
Sec.~\ref{sec:Pn}), and the $y$ axis lies perpendicular
to both the cusp direction and the observation direction, which
we can thus write
\beq
\ohat =  (\sin \theta, 0 ,\cos \theta)\,,
\eeq
where $\theta$ is the angle between cusp and observation directions.

We put the point of the cusp at $\sigma_\pm = 0$, and expand the string around that point,
\bml\label{eqn:X+-}
\beq\label{eqn:X+}
\bX_+ (\sigma_+) = \sigma_+ \hat \bz  + \frac12 \bx''_+ \sigma^2_+ + \frac16 \bx'''_+ \sigma^3_+
\eeq
and
\beq\label{eqn:X-}
\bX_- (\sigma_-) =  \sigma_- \hat \bz  + \frac12 \bx''_- \sigma^2_- + \frac16 \bx'''_- \sigma^3_-\,,
\eeq
\eml
where we defined $\bx''_\pm = \bX''_-(0)$ and $\bx'''_\pm =
\bX'''_-(0)$.  The constraints of the equations of motion require that
\beq\label{eqn:xppz}
{\bx_z^\pm}'' =0\,,
\eeq
and
\beq
{\bx_z^\pm}''' = - |\bx''_\pm|^2\,.
\eeq
We ignore other components of $\bx'''$, which contribute only at
higher orders in $\theta$, so Eqs.~(\ref{eqn:X+-}) become
\beq\label{eqn:X+-2}
\bX_\pm (\sigma_\pm) = \sigma_\pm \hat \bz + \frac12 \bx''_\pm \sigma^2_\pm - \frac16 |\bx''_\pm|^2 \sigma^3_\pm\hat \bz\,,
\eeq
and Eq.~(\ref{eqn:I}) can be written
\beq\label{eqn:I2}
\bI^\pm = \frac1l
\int_0^l \bX'_\pm e^{(i/2)(\omega \sigma_\pm - \bk\cdot \bX_\pm)} d\sigma_\pm\,,
\eeq
where $\omega$ is the frequency of the emitted radiation and
$k=\omega\ohat$ is its wavevector.

To describe the polarization of the gravitational waves, we need two
unit vectors lying in the plane perpendicular to $\ohat$.  We choose
\bea
\hat\bn_1 &=& \left(\cos \theta, 0,  -\sin \theta \right)\,,\\
\hat\bn_2 &=&\hat\by\,.
\eea
We use Eqs.~(\ref{eqn:dPdO},\ref{eqn:AA}) and convert from discrete to
continuous frequencies with $\omega=4\pi n/l$ to get the spectral
power density
\beq\label{eqn:P}
\frac{dP}{d\omega d\Omega}= \frac{G \mu^2 l^3 \omega^2}{8 \pi^2}
\left[|I^+_\perp|^2|I^-_\perp|^2+ 4\Im I^-_1 I^{-*}_2 \Im I^+_1 I^{+*}_2\right]\,,
\eeq
where $I^\pm_i = I^\pm(\bn_i) = \bn_i \cdot \bI^\pm$ and $|I^\pm_\perp|^2 = |I^\pm_1|^2+|I^\pm_2|^2$.

Since we are interested in the radiation near the cusp, we expand
$\bn_i$ and $\bk$ in the small parameter $\theta$ to get
\bea
\bk &=& \omega \left(\theta, 0 ,1 - \theta^2/2 \right)\\
\hat\bn_1 &=& \left(1 - \theta^2/2, 0, -\theta \right)\\
\hat\bn_2 &=&\left(0,1 , 0 \right)\,.
\eea
Using Eqs.~(\ref{eqn:X+-}),
\beq
I^\pm(\hat\bn) = \frac1l\int_0^l(\hat\bn_z + \sigma_\pm (\bx''_\pm \cdot \hat\bn))e^{(i\omega/4) \left( \theta^2 \sigma_\pm - (\bk \cdot \bx''_\pm) \sigma^2_\pm + |\bx''_\pm|^2 \sigma^3_\pm/3  \right)}  ~d\sigma_\pm\,.
\eeq
Using Eq.~(\ref{eqn:xppz}), the second derivatives of the $\bX^\pm$ at
the cusp can always be written
\beq
\bx''_- (\sigma_-) =  (\alpha_- \cos \phi_- ){\hat \bx} + (\alpha_- \sin \phi_- ){\hat \by} 
\eeq
and similarly,
\beq
\bx''_+ (\sigma_+) =  (\alpha_+ \cos \phi_+ ){\hat \bx} + (\alpha_+ \sin \phi_+ ){\hat \by}\,,
\eeq
giving
\beq\label{eqn:Ipm}
I^\pm(\hat\bn) = \frac1l\int_0^l (Q_\pm(\hat\bn) + P_\pm(\hat\bn) \sigma_\pm )~
e^{i \left( A_\pm \sigma_\pm + B_\pm \sigma^2_\pm + C_\pm \sigma^3_\pm  \right)}  ~d\sigma_\pm\,,
\eeq
where
\bea
Q_\pm(\hat\bn_1) &=&  \hat\bz \cdot \bn_1 = - \theta \\
P_\pm(\hat\bn_1) &=& \bx''_\pm \cdot \bn_1 = \alpha_\pm \cos \phi_\pm \\
Q_\pm(\hat\bn_2) &=&  \hat\bz \cdot \bn_2 = 0 \\
P_\pm(\hat\bn_2) &=& \bx''_\pm \cdot \bn_2 =
\alpha_\pm \sin \phi_\pm\\ 
A_\pm &=& \frac{\omega \theta^2}4 \\
B_\pm &=&  - \frac{\omega \theta}4 \alpha_\pm \cos \phi_\pm \\
C_\pm &=&  \frac\omega{12} \alpha_\pm^2\,,
\eea
and where we kept only the lowest order in $\theta$ in each term.
The integral in Eq.~(\ref{eqn:Ipm}) can be done in closed form,
giving 
\bea
I^\pm_1 = I^\pm(\hat\bn_1) &=& e^{-i \Phi_\pm} \left[\frac{2 }{\sqrt3}
  \frac{\theta^2}{\alpha_\pm l}  \sin^2  \phi_\pm \left( i \cos \phi_\pm  K_{2/3} (\xi_\pm) - |\sin \phi_\pm|  K_{1/3} (\xi_\pm) \right) \right]\\
I^\pm_2 =I^\pm(\hat\bn_2) &=& e^{-i \Phi_\pm}\left[ \frac{2 }{\sqrt3}
  \frac{\theta^2}{\alpha_\pm l} \sin \phi_\pm  |\sin \phi_\pm| \left( i  |\sin \phi_\pm| K_{2/3} (\xi_\pm) + \cos \phi_\pm  K_{1/3} (\xi_\pm) \right) \right]\,,
\eea
where $\Phi_\pm$ are some irrelevant phases,
\beq\label{eqn:xi}
\xi_\pm = \frac16 \omega \theta^3 \frac{|\sin^3 \phi_\pm|}{\alpha_\pm}\,,
\eeq
and $K$ is the modified Bessel function.  Thus
\bea
|I^\pm_1|^2+|I^\pm_2|^2
&=& \frac43 \frac{\theta^4}{\alpha_\pm^2 l^2}  \sin^4  \phi_\pm \left(K_{1/3} (\xi_\pm)^2 +K_{2/3}^2 (\xi_\pm) \right)\\ 
\Im I^\pm_1I^\pm_2
&=& \frac43  \frac{\theta^4}{\alpha_\pm^2 l^2}
\sin^4 \phi_\pm \sign(\sin\phi_\pm) K_{1/3}(\xi_\pm)K_{2/3} (\xi_\pm)\,.
\eea
Putting these in Eq.~(\ref{eqn:P}), we find
\bea\label{eqn:Pfinal}
\frac{dP}{d\omega d\Omega}&=& \frac{2G \mu^2 \omega^2\theta^8}{9 \pi^2l}\frac{\sin^4\phi_+\sin^4\phi_-}{\alpha_+^2\alpha_-^2} \bigg[\left(K^2_{1/3}(\xi_+) +K^2_{2/3}(\xi_+)\right)\left(K^2_{1/3}(\xi_-) +K^2_{2/3}(\xi_-)\right)\\
&&\qquad\qquad\qquad\qquad+4\sign(\sin\phi_+\sin\phi_-)K_{1/3}(\xi_+)K_{2/3}(\xi_+)K_{1/3}(\xi_-)K_{2/3}(\xi_-)\bigg]\,.\nonumber
\eea
Note that this expression gives the power emitted by the cusp per
frequency and per solid angle as a function of the the length of the
loop and four parameters that describe the cusp, namely $(\alpha_\pm,
\phi_\pm)$, which describe the crossing of the vectors $\bX'_+$ and
$\bX'_-$ on the Kibble-Turok sphere and the relative angle with respect
to the observation direction. Our code to compute the power spectrum
from individual loops in the simulation first looks at the possible
cusps in each loop and identifies these parameters.  We can then
integrate Eq.~(\ref{eqn:Pfinal}) over solid angle and over ranges of
frequency to include in the gravitational radiation spectrum from
triangles that are near cusps.

We do not use it for low frequencies where the approximation of the set of discrete harmonics by the
continuous frequency $\omega$ would lead to significant inaccuracy.

\section{The number of harmonics to compute}\label{sec:maxn}

\def\pomega{\varpi}             
\def\sigmamax{\sigma_{\text{max}}}
\def\rmax{r_{\text{max}}}

Except for directions near cusps, we find the power spectrum by
computing $I^{(n)}_\pm$ by fast Fourier transform.  This yields all
harmonics up through some maximum $\nmax$.  In most directions, the
power falls quickly, and we only need to compute a few harmonics.  But
in directions close to any $\bX'_\pm$, the corresponding $I^{(n)}_\pm$
may fall very slowly.  We estimate how many harmonics we need to
compute for any given direction as follows.

We consider the computation of $I^{+(n)}_x$ and suppress all $+$
subscripts and superscripts for this section.  We define
\bea
f(\sigma) &=& \sigma - X_z(\sigma)\\
g(\sigma) &=& X'_x (\sigma)\\
h(\sigma) &=& f'(\sigma) = 1 - X'_z(\sigma)
\eea
so that
\beq\label{eqn:maxnI1}
I^{(n)}_x = \frac1l\int_0^l d\sigma\, g(\sigma) e^{i\pomega_n f(\sigma)}\,,
\eeq
with $\pomega_n = 2\pi n/l$.  We can set the origin of coordinates
so that $\bx(0) = 0$.  Then as $\sigma$ goes from 0 to $l$,
$f(\sigma)$ also goes from 0 to $l$, and $h(\sigma)\ge 0$ so $f$ is
nondecreasing.

We can write Eq.~(\ref{eqn:maxnI1}) as a Fourier transform
\cite{Allen:1991bk}, by changing variables from $\sigma$ to $f$,
getting
\beq\label{eqn:maxnI}
I^{(n)}_x = \frac1l\int_0^l df\,s(f) e^{i\pomega_n f}\,,
\eeq
where
\beq
s(f) = \frac{g(\sigma(f))}{h(\sigma(f))}\,,
\eeq
and $\sigma(f)$ is the inverse of $f(\sigma)$.

We would like to bound $I_x$ by bounding the derivatives of $s(f)$.  We
integrate by parts $m$ times in Eq.~(\ref{eqn:maxnI}), finding
\beq
I^{(n)}_x = \frac{i^m}{l\pomega_n^m}\int_0^l df\,s^{(m)}(x) e^{i \pomega_n x}\,,
\eeq
where $s^{(m)}$ is the $m$th derivative of $s$.  If we can bound the
derivatives, $|s^{(m)}| <s^{(m)}_{\text{max}}$, then we will find
$|I^{(n)}_x|< B_m = s^{(m)}_{\text{max}}/\pomega_n^m$.

To differentiate $s(f)$, we can take $ds/df =
(ds/d\sigma)/(df/d\sigma)$.  The effect is to differentiate with
respect to $\sigma$ and then divide by $h$.  We thus have
\beq
s^{(m)} = \left(h^{-1}\frac{d}{d\sigma}\right)^m\left(h^{-1}g\right)\,.
\eeq
One term found in $s^{(m)}$ is the one where we repeatedly
differentiate the inverse power of $h$, which thus grows by two units
each step, giving
\beq\label{eqn:snh}
s^{(m)} \supseteq \frac{(2m-1)!! h'^m}{h^{-(2m+1)}}g\,.
\eeq
We conjecture that this is the dominant term.  Considering it alone,
we can derive a bound.  We need to know the largest value of
Eq.~(\ref{eqn:snh}) anywhere on the string.  Since we are not
concerning ourselves here with structure in $g$, we will merely
observe that $|g|< 1$.  Now $h = 1 - \cos\theta$, where $\theta$ is
the angle between $\bx'$ and the direction of observation.  We expect
that $|h'| = |X''_z|$ is not too large, because of smoothing.  Also
when $h$ is small, $\bX''$ is mostly transverse to the observation
direction.  So write
\beq\label{eqn:ssigma}
s^{(m)}(\sigma)\approx
(2m-1)!! \frac{r(\sigma)^m}{h(\sigma)}\,,
\eeq
where
\beq
r(\sigma) =\frac{|X''_z(\sigma)|}{h(\sigma)^2}\,.
\eeq
We're interested in $m\gg1$, so Eq.~(\ref{eqn:ssigma}) has its maximum
at the $\sigma$ that maximizes $r$, regardless of $m$.  Let us call
this point $\sigmamax$ and let $\rmax = r(\sigmamax)$.  Then
\beq
B_m = (2m-1)!!\frac{\rmax^m}{h(\sigmamax)\pomega_n^m}\,.
\eeq
Ignoring the extra power of $h$,
\beq
\ln B_m \approx m(\ln 2m - 1 +\ln(\rmax/\pomega_n))\,,
\eeq
which is minimized at $m=\pomega_n/(2\rmax)$, at which point
\beq
B_m \approx e^{-m} = e^{-\pomega_n/(2\rmax)}\,.
\eeq
Thus $I^{(n)}_x$ falls off as $e^{-\pomega_n/(2\rmax)}$.  For a given
string and a given $\ohat$, we scan the string to find $\rmax$ for
$I^+$ and $I^-$.  Using Eqs.~(\ref{eqn:dPdO},\ref{eqn:AA}) this gives
us an exponentially declining bound on $dP_n/d\Omega$ and thus a value
of $\nmax$ after which the power is insignificant.

When we do the calculation using this $\nmax$, we check that indeed
the computed $dP_n/d\Omega$ are small for the last few $n$.  Thus the
even if the conjecture above is not correct, we have good reason to
believe that we are not missing any power.

In certain cases, computational resources do not allow us to compute
as many harmonics as recommended above.  In particular, some loops
have ``pseudocusps'' \cite{Elghozi:2014kya,Stott:2016loe}, places
where $\bX'_+$ and $\bX'_-$ come close without crossing.  Because such
a point is not an actual cusp, it is handled by direct computation,
rather than our cusp code.  But in observation directions close to
$\bX'_+$ and $\bX'_-$, $h$ is very small and thus $\rmax$ large for
both $I^+$ and $I^-$, so the power remains high for many harmonics.
We limit the computation to $10^7$ harmonics, so we miss $n>10^7$
power coming from such regions.  This leads to a fictitious drop in
the computed power spectrum at $n = 10^7$, as shown in
Figs.~\ref{fig:PS-several-steps} and \ref{fig:PS-rad-mat}.  This is of
much less significance than it appears in the figures, because the
actual power is multiplied by $n^{-4/3}$ over what is shown there.

\section{Numerical summation of an infinite series}\label{sec:summation}

In this appendix we discuss the computation of an infinite sum,
\beq\label{eqn:sum1}
\sum_{n=1}^\infty A_n\,,
\eeq
such as appears in Eq.~(\ref{eqn:rhocp}).  Of course the $A_n$
must decrease rapidly enough so that the sum converges.  We can
evaluate only a finite number of $A_n$, and we must get from there to
an approximation for the infinite sum.

A great deal has been written on the subject of numerical integration,
but much less on numerical summation.  Most of what there is
involves the Euler-Maclaurin formula, which enables one to convert a
sum of discrete samples of a smooth function into an integral.  But
here we do not have samples of a smooth function but rather a function
defined only at discrete values $n$.  Thus we will develop a little of
the needed techniques for numerical summation by analogy with
numerical integration.

A numerical integration method can be though of as a way of taking a
finite number of samples of the integrand, producing from those an
approximation to the integrand, and integrating that instead.  For
example, in the trapezoidal rule, the integrand is approximated by
linear interpolation between sampled points.  We will use a similar
technique here.

The standard procedure for integrals going to infinity (for example
see Ref.~\cite{Press:NR}) is to perform a change of variable to render
the integration range finite.  For example if one has
\beq\label{eqn:int1}
\int_1^\infty dx\,f(x)\,,
\eeq
one can let $t=1/x$ to get 
\beq
\int_0^1 dt\,f(1/t)/t^2\,.
\eeq
If $f(x)$ decreases at least as fast as $1/x^2$ \cite{Press:NR}, then
$f(t)/t^2$ will be bounded as $t\to0$.

In our case, $A_n = C_n P_n$.  For very large $n$, the power $P_n$ is
dominated by cusp emission and goes as $n^{-4/3}$.  The coefficients
$C_n$ decrease, so $A_n$ decreases at least as $n^{-4/3}$ but not
necessarily faster.  If we had $f(x)$ going as $x^{-4/3}$ in
Eq.~(\ref{eqn:int1}), we should change variables to $t=x^{-1/3}$,
giving
\beq\label{eqn:int2}
3\int_0^1 dt\,f\left(t^{-3}\right)/t^4\,,
\eeq
where the integrand is bounded as $t\to0$.

We will now use Eq.~(\ref{eqn:int2}) as a guide to approximate
Eq.~(\ref{eqn:sum1}) using a finite number of $n$.  The discrete
approximation to the integrand in Eq.~(\ref{eqn:int2}) is $S_n =
n^{4/3} A_n$, and it is this $S_n$ that we will interpolate between
computed values.  Furthermore the variable of interpolation, analogous
to $t$, should be $n^{-1/3}$.  Thus if we have computed $S_n$ and
$S_m$, we will find $S_l$ for $l\in(m,n)$ by
\beq\label{eqn:interpolateS}
S_l = \frac{m^{-1/3}-l^{-1/3}}{m^{-1/3}-n^{-1/3}}S_n
+ \frac{l^{-1/3}-n^{-1/3}}{m^{-1/3}-n^{-1/3}}S_m\,.
\eeq
The sum of terms from $m$ through $n-1$ is given by
\bea\label{eqn:sum2}
\sum_{l=m}^{n-1} n^{-4/3} S_l
=&& \frac{S_m - S_n}{m^{-1/3}-n^{-1/3}}[\zeta(5/3,m)-\zeta(5/3,n)]\\
&&-\frac{n^{-1/3}S_m - m^{-1/3}S_n }{m^{-1/3}-n^{-1/3}}[\zeta(4/3,m)-\zeta(4/3,n)]\,,
\nonumber
\eea
where
\beq
\zeta(s, m) = \sum_{k=0}^\infty (m+k)^{-s}
\eeq
is the Hurwitz $\zeta$ function.

Suppose we have computed $S_n$ for some set of $n_j$, $j = 1\ldots
N$.  For simplicity, let us require that $n_N = \infty$.  Of course
$A_\infty = 0$, but if $C_n$ approaches a nonzero limit as $n\to\infty$,
then $S_\infty$ is a constant that we can compute, and using it
improves the approximation.  This occurs when we compute the
total power $\Gamma$, where $C_n = 1$.

We can write
\beq
\sum_{n=1}^\infty A_n=\sum_{n=1}^\infty n^{-4/3} S_n=
\sum_{j=1}^{N-1} \sum_{l=n_j}^{n_{j+1}-1} l^{-4/3} S_l
\approx \sum_{j=1}^N c_j S_{n_j}\,,
\eeq
where $c_j$ is the sum of the coefficient of $S_m$ in
Eq.~(\ref{eqn:sum2}) with $m=n_j$, $n=n_{j+1}$ and the coefficient of
$S_n$ in Eq.~(\ref{eqn:sum2}) with $m=n_{j-1}$, $n=n_j$,
\bea\label{eqn:cj}
c_j =&&\frac{\zeta(5/3, n_j) - \zeta(5/3, n_{j+1})
- [\zeta(4/3, n_j) - \zeta(4/3, n_{j+1})]n_{j+1}^{-1/3}}
{n_j^{-1/3} - n_{j+1}^{-1/3}}\\
&-&\frac{\zeta(5/3, n_{j-1}) - \zeta(5/3, n_j)
- [\zeta(4/3, n_{j-1}) - \zeta(4/3, n_j)]n_{j-1}^{-1/3}}
{n_{j-1}^{-1/3} - n_j^{-1/3}}\,.
\nonumber
\eea
For $j = 1$, there is no contribution from the previous interval.  For
$j=N$, there is no contribution from the next interval.

Equation~(\ref{eqn:interpolateS}) still holds with $n=\infty$ and
consequently $n^{-1/3} = 0$.  Then Eq.~(\ref{eqn:cj}) holds also.  For
$c_{N-1}$, everything vanishes in the first line except the first terms
in the numerator and the denominator, while the second line is normal.
For $c_N$ the first line is absent because it is the last interval, and
in the second line all terms involving $n_j$ vanish.

One might approximate an integral such as Eq.~(\ref{eqn:int2}) by
evaluating the integrand at evenly spaced $t$.  By analogy, we can
choose the $n_j$ so that the $n_j^{-1/3}$ are evenly spaced, as much
as possible.  We do this by picking a fiducial number $N'$, in our
case 1000, choosing $t_i = i/N'$ for $i=0\ldots N'$, and letting
$\{n_j\}$ be the distinct integers, plus infinity, found by rounding
the $t_i^{-3}$.  The number of such modes is about $4
(N'/3)^{3/4}$.  In our case $N=312$.

\section{Handling of rare bursts}\label{sec:bursts}

The above computation of $\ogw$ is the computation of its average
value.  If what we observe is the total contribution due to many
loops, then by the central limit theorem we should expect a Gaussian
background.  But if the average is dominated by a few rare
bursts, so rare that we might not have seen any of them, then we
should expect a smaller signal.  Thus rare bursts should be excluded
from the background calculation \cite{Damour:2001bk}.

Suppose an experiment runs for time $T$ and reports the average signal
at some typical frequency $f$.  If strong bursts occur less often than
the duration of the experiment, then we would probably not have seen
even one, so their contribution should be excluded from our estimate
of the average power.  So the question is whether any significant
contribution to $\ogw$ comes from strong bursts that typically do not
occur within a time interval $T$.  We will see below that it does not.

One can also consider the status of bursts that occur with frequency
greater than $1/T$ but less than frequency $f$.  This might matter for
experiments such as LIGO and LISA, but not for pulsar timing, where
the typical frequency is about the inverse of the observation time.
Bursts with these intermediate rates contribute to the average power
over the entire interval $T$ in the usual way, so if that is the
observation with which we compare, they do not need to be excluded.
In fact, such bursts are likely to be detected by burst detection
pipelines, rather than being reported as part of the background.
However, this makes the effect more detectable, not less.  So
including intermediate-rate bursts makes no mistake in detection
through the average power, but neglects the possibility of detection
of bursts as bursts.  That, however, is the subject of a different
body of work, and here we will show that there is no need to exclude
burst with rates less than $1/T$.

We will show that rare bursts are not important by analyzing a
particular population of bursts that are stronger than those that make
significant contributions to the background and nevertheless occur
frequently in period $T$.  There are two factors that lead to an
energetic burst: large loop length $l$, and recent emission, i.e.,
small redshift.

We are concerned with tightly beamed bursts emitted by cusps, which
means with radiation at frequencies high compared with the loop
oscillation frequency $2/l$.  Thus the discrete nature of the loop
harmonics is not relevant, and we can consider a continuous form of
the power, $P(y)$, with $y=(1+z)fl$, where $f$ is the observed
frequency today.  We define $P(y)$ to be the power per unit $y$ from
the given loop, so the power per unit range of observed frequency is
$P(y)dy/df = (1+z)l P(y)$.

We now compute the dependence of the burst energy density on $l$ and
$z$.  The period in the emitting frame between burst emissions is
proportional to $l$.  Thus we multiply by $l$ to convert the power
emitted into the energy permitted per burst.  Then we divide by $1+z$
because the energy is decreased by redshifting.  Thus the present-day
energy of a burst per unit range of observed frequency is proportional
to $l^2 P(y)$.  Since $P(y)\propto y^{-4/3}$ \cite{Vachaspati:1984gt}, this goes as
$l^{2/3} (1+z)^{-4/3}$.

Now we consider beaming.  We define an approximate beaming angle
$\theta$ by setting $\xi_\pm=1$ in Eq.~(\ref{eqn:xi}).  Since
$\alpha_\pm \sim 1/l$, we find $\theta \sim (l\omega)^{-1/3} \sim
y^{-1/3}$.  Beaming thus enhances the burst energy density by a factor
of $\theta^2 \sim y^{2/3}$, giving in all $l^{4/3} (1+z)^{-2/3}$.

In addition, the energy of the burst is diluted by the square of the
proper distance to the point of emission, which is proportional to $z$
for $z\ll 1$ and asymptotes to the horizon distance for large $z$.
Again, recent bursts are stronger.

Since recent bursts from large loops are the strongest, we will
consider bursts coming from strings of length around some
specific $l > \Gamma G \mu t_0$ at places with $z<1$.  Even
these large loops are dominated by radiation-era relics, so we can use
Eq.~(\ref{eqn:nrG}), with $G_1=1$, $z\ll 1$, and $l>\Gamma G \mu t_0$,
\beq
\nsf_r(l,t_0) \approx  \frac{0.5 (H_r^2)^{3/4}}{l^{5/2}}\,.
\eeq
Using Eq.~(\ref{eqn:Omegar}), we find
the number of loops per logarithmic interval in $l$,
\beq
l \nsf_r(l,t_0) \approx 9.3 \times 10^9 \left(\frac l{\year}\right)^{-3/2}
  \Gpc^{-3}\,.
\eeq

The proper distance to $z=1$ is about $3.3$ Gpc \cite{Wright:2006up},
so the volume is $150 \Gpc^3$, and the total number of loops
\beq
N(\ln l) \approx 1.4 \times 10^{12} \left(\frac l{\year}\right)^{-3/2}\,.
\eeq
Almost all our smoothed loops have 2 cusps per oscillation, so each
loop produces bursts at rate $4/l$.  The fraction of bursts we can see
is given by the fraction of solid angle occupied by the beam,
$\theta^2/4$, where, as before, $\theta \approx (fl)^{-1/3}$.  Thus
the rate of bursts received from a population of loops with lengths
around $l$ and $z<1$ is
\beq\label{eqn:R}
R = 1.4 \times 10^{12} \left(\frac l{\year}\right)^{-19/6}
\left(f\cdot\year\right)^{-2/3}\year
= 1.4 \times 10^{7} \left(\frac l{\year}\right)^{-19/6}
\left(\frac{f}{\Hz}\right)^{-2/3}\year\,.
\eeq

For pulsar timing, we take $f = (5\year)^{-1}$ and consider loops
around $l=1000\year$.  Then we find about $4\times 10^7$ loops
emitting $2\times 10^5$ bursts per year, of which we can see about
0.7\%, giving 1400 bursts per year.  Even these large loops give
frequent events which would be seen as a Gaussian background.  

We found \cite{Blanco-Pillado:2017rnf} that $G \mu < 2 \times 10^{-11}$, so
$\Gamma G \mu < 10^{-9}$ and the dominant size of loops today, $\Gamma
G \mu t_0$, is no more than $14 \year$.  Thus the loops we just
considered with $l\sim 1000\year$ contribute a negligible fraction of
the total background, and ignoring loops larger than these has no
effect.

Turning now to LISA, we choose $f = 10^{-2}$ Hz and consider loops
around $l=100\year$, still several times larger than $\Gamma G \mu
t_0$.  Then Eq.~(\ref{eqn:R}) gives $R=140/\year$, so once again rare
bursts do not need to be excluded.

For LIGO, we choose $f = 10^2$ Hz and consider loops around
$l=14\year$, finding $R=150/\year$.  Thus loops right at $\Gamma G \mu
t_0$ do not need to be excluded, but significantly larger ones might.
But this is very far from making a difference to the background seen
by LIGO, which comes almost entirely from emission during the
radiation era.  Even excluding all matter-era emission would make
little difference.

\bibliography{no-slac,stochastic-gw-from-strings}

\end{document}